\documentclass[epj]{webofc}
\usepackage[varg]{txfonts}
\usepackage{graphicx}

\wocname{EPJ Web of Conferences}

\woctitle{CONF12}

\newcommand{\vect}[1]{\mathbf{#1}}
\newcommand{\gvect}{\boldsymbol}

\newcommand*\colvec[3][]{
    \begin{pmatrix}\ifx\relax#1\relax\else#1\\\fi#2\\#3\end{pmatrix}
}

\newcommand{{\SD}}{\rm SD}

\newcommand{{\Mc}}{\mathcal{M}}

\newcommand{\ver}{\mbox{\boldmath${\rm r}$}}
\newcommand{\vesig}{\mbox{\boldmath${\rm \sigma}$}}

\newcommand{\veP}{\mbox{\boldmath${\rm P}$}}
\newcommand{\vep}{\mbox{\boldmath${\rm p}$}}

 \newcommand{\veA}{\mbox{\boldmath${\rm A}$}}

\newcommand{\veB}{\mbox{\boldmath${\rm B}$}}

\newcommand{\veF}{\mbox{\boldmath${\rm F}$}}

\newcommand{\lan}{\langle}
\newcommand{\ran}{\rangle}

\newcommand{\be}{\begin{equation}}
\newcommand{\ee}{\end{equation}}

\newcommand{\ben}{\begin{equation*}}
\newcommand{\een}{\end{equation*}}

\newcommand{\bea}{\begin{eqnarray}}
\newcommand{\eea}{\end{eqnarray}}

\begin{document}
\selectlanguage{english}

\title{Mesons in ultra-intense magnetic field: an evaded collapse}

\author{B.O. Kerbikov \inst{1,2,3}\fnsep\thanks{\email{borisk@itep.ru}} \and M.A. Andreichikov\inst{1}\fnsep\thanks{\email{andreichicov@mail.ru}} \and Yu.A. Simonov\inst{1}\fnsep\thanks{\email{simonov@itep.ru}}}

\institute{A.I.~Alikhanov Institute for Theoretical and Experimental Physics,
Moscow 117218, Russia
\and
 Lebedev Physical Institute, Moscow 119991, Russia          
\and
Moscow Institute of Physics and Technology, Dolgoprudny 141700,
Moscow Region, Russia         
}

\abstract{
Spectra of $q\bar q$ mesons are investigated in the framework of the
Hamiltonian obtained  from the relativistic path integral in external
homogeneous magnetic field. The spectra of all 12 spin-isospin $s$-wave states
generated by $\pi$- and $\rho $-mesons with different spin projections, are
studied analytically as functions of the field strength. Three types of
behavior with characteristic splittings are found. The results are in agreement
with recent lattice calculations.}

\maketitle

\section{Introduction \label{intro} }

The interest  to  the  behavior of quarks, hadrons and atoms in strong magnetic
field (MF) has been very high  during the last decade. The outbrake of the
research   activity in  this field was inspired by the fact that MF up to $eB
\sim \Lambda^2_{QCD} \sim 10^{19} G$ \footnote{ We use the relativistic system
of units $\hbar =c=1, e^2=4 \pi \alpha$. Then 1 GeV$^2 \simeq 5.12\cdot 10^{19}
G$} is generated during the early stages of peripheral heavy-ion collisions at
RHIC and LHC. The field about four orders of magnitudes less is  anticipated to
operate in magnetars. The immediate question is what happens to the mass and
the wave function of a meson embedded in such a strong MF. The answer to this
question has been searched for in various approaches (see \cite{1} for a list
of references) including lattice simulations. In  the present work the problem
is investigated in the framework of the relativistic path integral Hamiltonian
(PIH) formalism \cite{2,3,4}. For pion this method has to be supplemented by
the elements of chiral dynamics \cite{5}. The analytical results will be
compared with the lattice calculations presented recently in \cite{1}. Before
getting involved with the details of calculations it makes sense to relate the
MF strength to some characteristic physical parameter which defines the
spectrum of quark-antiquark meson states. From the  textbooks we know that for
the hydrogen atom the critical, or the so-called ``atomic field'', is $B_a =
\frac{\alpha^2m^2_e}{|e|} = 2.35\cdot 10^{9} G$. This value corresponds to the
situation when the magnetic, or Landau radius $l_B = (|e|B)^{-1/2}$ is equal to
the Bohr radius. The QCD coupling constant $\alpha_s\sim 1$, the meson radius
at $eB =0$ is determined by the QCD string tension $\sigma \simeq (0.15 -0.18)
$ GeV$^2$ \cite{4}. It is therefore natural to define for the hadron spectra
the critical MF as $B_\sigma = \sigma/|e|\simeq 10^{19} G$ which yields $l_B
\simeq 0.6 $ fm. This value is  approximately equal or smaller than the typical
hadron size.

The determination of the hadron spectrum in MF is not an easy task. The first
problem is to separate the center-of-mass (c.m.) motion. For the neutral
nonrelativistic system in MF this can be done (with some qualifications) making
use of the Pseudomomentum \cite{6,7,8,9}. This approach was extended to the
relativistic sector within the PIH framework in \cite{4}. For a charged meson
Pseudomementum method is  applicable only for an unphysical model of a meson
with two equally charged quarks \cite{4}. In this contribution we present the
results on the meson spectrum in MF both within the Pseudomomentum approach and
in a new analytical method of  constituents separation (CS). It will be argued
that its accuracy is  within 15\% for ultra-strong MF $(eB\gg \sigma)$, and
within 20\% for $eB < \sigma$. The  method allows to study neutral and charged
mesons in the same way. The results for the neutral mesons will be obtained
both in Pseudomomentum and CS approaches. In  this way the accuracy of the CS
will be tested.

The most important question we  have to answer is whether the meson spectrum in
MF is bounded from below. In other words, does the meson mass reaches zero
value at some MF strength. We shall point out the two dynamical mechanisms that
might have led to  such a collapse and explain why this does not happen.
 The paper is organized as follows. In section 2 the relativistic Hamiltonian
 based on path integral Feynman-Fock-Schwinger representation is written down
 and  the spectral problem is formulated. In section 3 we discuss the possible
 types of meson mass trajectories in MF. Section 4 contains the analysis of
 perturbative corrections and potential reasons for the collapse of meson state
 in MF. In  section 5 we present the main results in comparison with lattice
 calculations.

\section{The relativistic Hamiltonian and the spectral problem}

To find the meson masses in MF we use  the path integral Hamiltonian (PIH)
method  based  on the Feynman-Fock-Schwinger representation \cite{2,3,4}. It
allows  with the help of Wilson loop to treat the interaction of quarks with
external Abelian and non-Abelian fields in a gauge-invariant way. As it was
shown in \cite{2,3,4,10} the  quark-antiquark spectral problem in MF in PIH
formalism is reduced to the bound states problem for the relativistic
Hamiltonian which includes all the  non-perturbative dynamics\be H_{q\bar q} =
\sum^2_{i=1} \frac{(\vep^2_i - e_i \veA_i)^2 + m^2_i +\omega^2_i - e_i
(\vesig_i \veB)}{2\omega_i} + \sigma |\ver_1-\ver_2|.\label{1}\ee

Here $\omega_i$ is the $i$-th quark dynamical mass, or the einbein variable
\cite{2,3,4}. The MF is convenient to  take  in the  symmetric gauge
$\veA_i=\frac12 \left(\veB\times \ver_i\right)$ since this gauge allows to
define the angular momentum projection of each quark as a quantum number. The
next step   is to perform minimization with respect to $\omega_i$ which yields
the physical spectrum \be H_{q\bar q} \psi_n = M_n \psi_n,\label{2}\ee \be
\frac{\partial M_n}{\partial\omega_i} =0.\label{3}\ee The total meson mass is a
sum of the non-perturbative (dynamical) one obtained from (\ref{1})-(\ref{3})
and the first-order perturbative contributions \be M_{tot} = M_0+ \lan \psi_0
|v_{\rm oge}| \psi_0 \ran + \lan a_{ss} \ran (\vesig_1\vesig_2) + \delta
M_{SE},\label{4}\ee where $V_{\rm oge}$ is the one-gluon exchange potential,
and  $a_{ss}$ and $\Delta M_{SE}$ are the spin-spin and self energy
contributions.

For the neutral hadrons (mesons and baryons) the eigenvalue problem (\ref{2})
admits the  exact solution which is obtained by the separation of the c.m.
motion. To this end the pseudomomentum operator is introduced \cite{6,7,8,9}
\be \hat\veF = \sum^2_{i=1} \left[ \veP_i + \frac12 e_i (\veB\times
\ver_i)\right].\label{5}\ee

In MF the pseudomomentum takes the role of the mechanical momentum, commutes
with the Hamiltonian, and  is therefore a  constant of motion. Physically,
$\veF$ is conserved since it takes into account the Lorentz force acting on
particles in MF.

In \cite{1} we have proposed a more general approach which allows to
investigate the mass spectra of both neutral and charged mesons. This is the
constituent separation (CS) method. The c.m. position $\ver_0$ is fixed at the
origin and an effective string tension $\sigma_i$ is attributed to each quark
\be \sigma |\ver_1-\ver_2| \to \sigma_1 |\ver_1-\ver_0| +\sigma_2
|\ver_2-\ver_0| .\label{6}\ee In this picture quarks may be considered  as
quasi-independent of the non-perturbative part of the interaction.

\section{Meson trajectories in strong MF}

In strong MF $eB \gg \sigma$ it is convenient to use for the spin degrees of freedom in (\ref{1}) a basis in which the operator $\vect{B}(\frac{e_1}{2\omega_1} \gvect{\sigma}_1 + \frac{e_2}{2\omega_2} \gvect{\sigma}_2)$ is diagonal. The four vectors forming this basis are $|++ \rangle,\ |-- \rangle,\ |+- \rangle,\ |-+ \rangle$. One can easily obtain three types of asymptotic meson trajectories at $eB \rightarrow \infty$. The character of the trajectory is determined by the signs of the quark charges and the spin directions. According to the terminology adopted in atomic physics the trajectory is called low-field seeking (LFS) if the energy decreases as the MF decreases. The state which at $eB \rightarrow \infty$ is MF-independent may be called zero-field seeking (ZFS). The types of asymptotic trajectories are
\begin{multline}
\label{7}
 \begin{split}
  a)&\ ZFS:\ e_1 \sigma_1^{z} >0,\ e_2 \sigma_2^{z} >0 :\ M_{ZFS}(eB \rightarrow \infty) = 2 \sqrt{\sigma},\\
  b)&\ LFS1:\ e_1 \sigma_1^{z} >0,\ e_2 \sigma_2^{z} <0 :\ M_{LFS1}(eB \rightarrow \infty) = \sqrt{2e_1B} + \sqrt{2\sigma},\\
  c)&\ LFS2:\ e_1 \sigma_1^{z} <0,\ e_2 \sigma_2^{z} <0 :\ M_{LFS2}(eB \rightarrow \infty) = \sqrt{2e_1B} +\sqrt{2e_2B} ,\\
  \end{split}
\end{multline}
The trajectory LFS2 exhibits stronger MF dependence than LFS1. From (\ref{7}) it follows that the $(\pi^+,\ \rho^+)$ family which contains $u$ and $\bar d$ quarks is distributed among the three above classes in the following way: $\rho^+(s_z = 1)$ belongs to ZFS, $\pi^+(s_z=0)$ and $\rho^+(s_z=0)$ belong to LFS1, $\rho^+(s_z=-1)$ rests in LFS2. The same situation up to the sign change holds for $(\pi^-,\ \rho^-)$. The states $\pi^0$ and $\rho^0(s_z=0)$ contain $u \bar u$ and $d \bar d$ components. The charges of $u$ and $d$ are different and this results in an additional double splitting.

\section{Perturbative  corrections}

\begin{figure}[h]
 \centering
    \includegraphics[width=.6\textwidth]{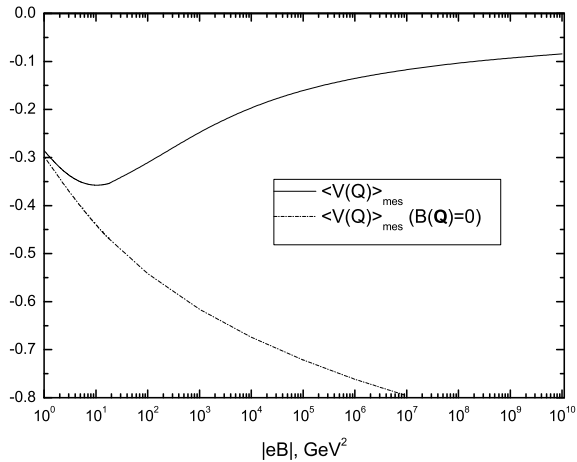}
    \caption{Color Coulomb matrix element $\langle \psi_0 | V_{OGE}| \psi_0 \rangle$ (in GeV) in MF with  screening (solid line) and without screening (dashed line) by $q \bar q$ pairs}
\label{fig1}
\end{figure}
As it was shown in \cite{10}, the color Coulomb interaction in presence of the MF contains a potential danger of a collapse for ZFS states. This reminds the ''fall-to-the-center'' phenomenon in the hydrogen atom enbedded in strong MF. The one-gluon exchange matrix element $\langle \psi_0 | V_{OGE}| \psi_0 \rangle$ has a negative sign, its absolute value grows with $eB$, and as shown on Fig.\ref{fig1} ''fall-to-the-center'' takes place at $eB \sim 10 \ GeV^2$. To prevent the collapse the screening by the quark loops at the lowest Landau level (LLL) was introduced in \cite{10}. 

As one can see from Fig.\ref{fig1}, the color Coulomb collapse is really evaded. We remind that in superstrong MF radiative corrections screen the Coulomb potential in the hydrogen atom thus leading to the freezing of the ground state energy at the value $E_0 = -1.7 \ KeV$ \cite{11, 12}. It is interesting to note that asymptotically at $eB \rightarrow \infty$ the matrix element $\langle \psi_0 | V_{OGE}| \psi_0 \rangle$ vanishes.

Another threat of a collapse comes from the hyperfine spin-spin interaction $\langle a_{SS} \rangle$. In the first-order perturbation theory in PIH formalism it corresponds to the color-magnetic interaction of the form 
\begin{equation}
  \label{10}
  V_{SS} = \frac{8\pi \alpha_s^{(0)}}{9 \omega_1 \omega_2} \delta(\vect{r}_1 - \vect{r}_2) \gvect{\sigma}_1 \cdot \gvect{\sigma}_2
\end{equation}
In strong MF the ground state wave function acquires the form of an ellipsoid elongated in the direction of MF. At $eB \rightarrow \infty$  the transverse and longitudinal radii are $r_{\perp} \sim 1/\sqrt{eB}$, $r_z \sim 1/\sqrt{\sigma}$. This means that the focusing of the wave function at the origin and a divergent factor $|\psi(0)|^2 \sim eB$ in the matrix element of $V_{SS}$. Not that the problem of singularity due to $\delta$-function interaction exists without MF as well. It is cured by smearing $\delta$-function \cite{13,14}. In PIH formalism there is a natural cut-off parameter $\lambda \sim 1 \ GeV^{-1}$. It corresponds to the correlation length of the stochastic vacuum gluonic field. The $\delta$-function is replaced by 
\begin{equation}
  \label{11}
  \delta(\vect{r}) \rightarrow \frac{1}{\pi^{3/2} \lambda^3}e^{-\frac{\vect{r}^2}{\lambda^2}}.
\end{equation}
In this way the ''fall-to-the-center''is prevented for all ZFS states except for the $\pi^0$-meson. The $\pi^0$ trajectory is stabilized if one takes its chiral degrees of freedom into account. We also note that in \cite{3} a general theorem was proven according to which the eigenvalues of the relativistic Hamiltonian in MF are positive. The explicit account of pion chiral dynamics \cite{5} confirm this result.  The main point is that GMOR relations remain valid for neutral pions in arbitrary-strong MF, while charged pions loose their chiral properties at $eB > \sigma$.

\section{Results and conclusions}

\begin{figure}[h]
\centering
    \includegraphics[width=.6\textwidth]{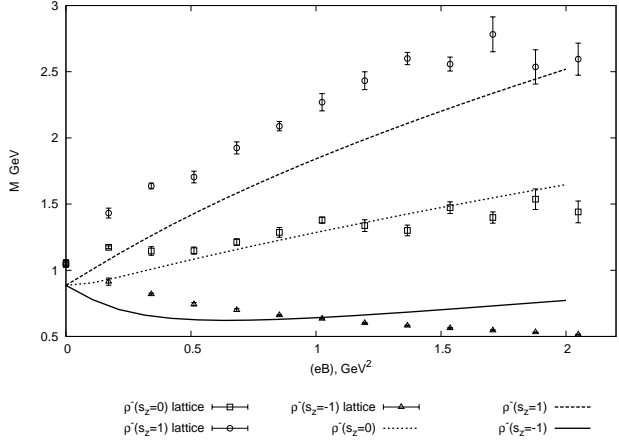}
    \caption{The $\rho^{-}$ meson mass evolution in MF from our analytic and lattice data.}
\label{fig2}
  \end{figure}
Below we present the results of our analytic calculations in comparison with the recent lattice results from \cite{1}. In Fig.\ref{fig2} the $\rho^-$ meson mass evolution in MF is shown.

\begin{figure}[h]
\centering
    \includegraphics[width=.6\textwidth]{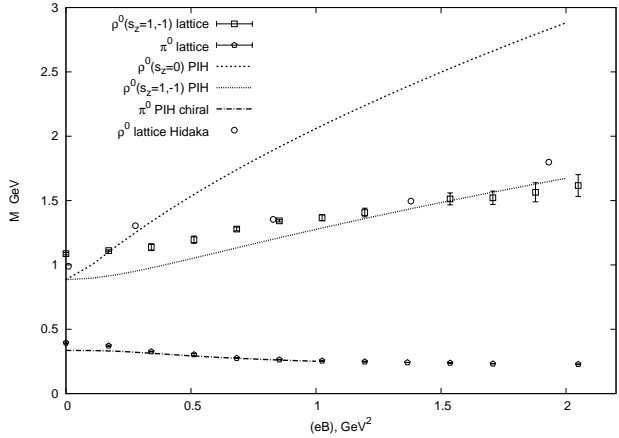}
    \caption{Mass evolution of $(\pi^0,\ \rho^0)(u \bar u)$ family from analytic and lattice data (hollow circles are from \cite{17}).  }
\label{fig3}
  \end{figure}
\begin{figure}[h]
\centering
    \includegraphics[width=.6\textwidth]{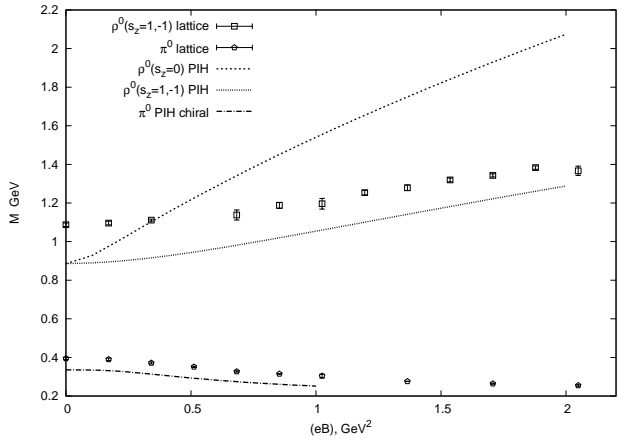}
    \caption{Mass evolution of $(\pi^0,\ \rho^0)(d \bar d)$ family from analytic and lattice data  }
\label{fig4}
  \end{figure}
\begin{figure}[h]
\centering
    \includegraphics[width=.6\textwidth]{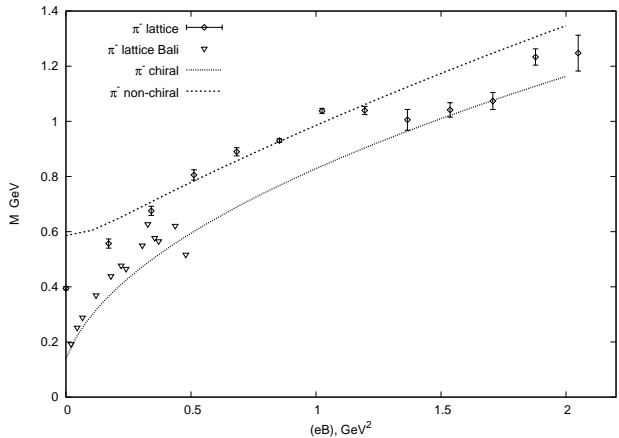}
    \caption{Mass evolution of the chiral (solid line) and non-chiral (dashed line) $\pi^-$ meson in comparison with lattice data (triangles are from \cite{bali}.}
\label{fig5}
  \end{figure}
In Fig.\ref{fig3} and Fig.\ref{fig4} results for $\pi^0$ and $\rho^0$ are exposed. One should keep in mind that $u \bar u$ and $d \bar d$ components give rise to their own trajectories. The growing trajectories belong to the LFS2 class and the splitting is equal to $\sqrt{2}$. In Fig.\ref{fig5} we present the mass evolution of chiral and non-chiral $\pi^-$ in comparison with the lattice data. The chiral effects provide the decrease of the mass to its physical value at $eB \rightarrow 0$.

In this work we have evaluated the trajectories of $\pi$ and $\rho$ meson masses as functions of the external MF. The meson quark content and pion chiral dynamics were thoroughly taken into account. The most interesting problem was whether the mass remains finite in arbitrary strong MF. The collapse might have happened either due to color Coulomb interaction, or due to spin-spin potential proportional to $\delta$-function. We have shown that in both cases there are physical reasons why collapse is evaded. The analytic calculations of meson mass trajectories give the results which are in agreement with recent lattice simulations.

 \vspace{0.5cm}
 \noindent
 {\bf Acknowledgement}\\ The authors are  grateful to E.V.Luschevskaya and O.E.Solovjeva for  kindly providings with their lattice results and to O.V.Teryaev for discussions. The authors are  supported by the Russian Science Foundation grant number 16-12-10414.

\end{document}